%% file: cas_gibbon.tex
\documentclass{cernrep}

\graphicspath{{figures/}}

\input macros
\usepackage[strings]{underscore}
\usepackage[T1]{fontenc}
\usepackage{color}
\usepackage[normalem]{ulem}
\usepackage{url}

\DeclareUrlCommand\ULurl{%
  \renewcommand\UrlLeft{\uline\bgroup}%
  \renewcommand\UrlRight{\egroup}}
\usepackage{fancyhdr}
\fancyhfoffset{4 mm}
\fancypagestyle{ARTTITLE}{%
\fancyhf{} 
\lhead{\footnotesize{Proceedings of the 2019 CERN--Accelerator--School course on
\it{ High Gradient Wakefield Accelerators}, Sesimbra, (Portugal)}}
\lfoot{Available online at \ULurl{https://cas.web.cern.ch/previous-schools}}
\rfoot{\thepage\hspace*{3mm}}

   }
\begin{document}
\title{Introduction to Plasma Physics}
\author{Paul Gibbon}
\institute{Forschungszentrum J\"ulich GmbH, Institute for Advanced Simulation, J\"ulich Supercomputing Centre, D-52425 J\"ulich, Germany \\ 
Centre for Mathematical Plasma Astrophysics, Katholieke Universiteit Leuven, 3000 Leuven, Belgium}

\begin{abstract}
The following notes are intended to provide a brief primer in plasma physics, introducing  common definitions, basic properties and processes typically found in plasmas. These concepts are inherent in contemporary plasma-based accelerator schemes, and thus build foundation for the more advanced lectures which follow in this volume. No prior knowledge of plasma physics is required, but the reader is assumed to be familiar with basic electrodynamics and fluid mechanics.
\end{abstract}
\keywords{Plasma definitions; wave propagation; electron motion; ponderomotive force.}
\maketitle
\thispagestyle{ARTTITLE}
\section{Plasma types and definitions}

Plasmas are often described as the fourth state of matter, alongside gases, liquids and solids; a definition which
does little to illuminate their main physical attributes. In fact, plasmas can exhibit behaviour characteristic of all three
of the more familiar states depending on its density and temperature, so we obviously need to look for other distinguishing features.  A simple textbook definition \cite{chen:book,dendy:book} would be: a~\textit{quasi-neutral} gas of charged particles showing \textit{collective behaviour}, which sounds a bit more authoritative, but demands further explanation of the rather fuzzy-sounding `quasi-neutrality' and `collectivity'.  The first of these is actually just a mathematical way of saying that
even though the plasma particles consist of freely moving electrons and ions, their overall charge densities cancel each other in equilibrium.  So, if $n_e$ and $n_i$ are the number densities of electrons and ions with charge state $Z$  respectively,
then these are \textit{locally balanced}:
\begin{equation}
n_e \simeq Z n_i.
\label{quasineut}
\end{equation}

The second property, collective behaviour, arises because of the long range nature of the $1/r$ Coulomb potential, which means that local disturbances in equilibrium can have a strong influence on remote regions of the plasma. 
In other words, macroscopic fields usually dominate over short-lived microscopic fluctuations, and a net charge imbalance $\rho=e(Zn_i-n_e)$ will immediately give rise to an~electrostatic field according to Gauss' law:
$$ \nabla.\bm{E} = \frac{\rho}{\varepsilon_0}.$$ 
Likewise,  the same set of charges moving with velocities $v_e$ and $v_i$ respectively, will give rise to a \textit{current} density $J=e(Zn_iv_i-n_ev_e)$. This in turn induces a magnetic field according to Amp\`eres law:
$$ \nabla \cross \bm{B} = \mu_0 \bm{J}.$$
It is these internally driven electric and magnetic fields which largely determine the dynamics of the~plasma, including its response to externally applied fields through particle or laser beams -- like, for example, in the case of plasma-based accelerator schemes.

Now that we have established what plasmas are, it is natural to ask where we can find them.  In fact they are rather ubiquitous: in the cosmos, 99\% of the visible universe is in a plasma state: stars, the~interstellar medium and jets of material emanating from various astrophysical objects. Closer to home, the~ionosphere extending from around 50 km = 10 Earth-radii to 1000 km provides vital protection to life on Earth from solar radiation. Terrestrial plasmas can be found in fusion devices, machines designed to confine, ignite and ultimately extract useful energy from deuterium-tritium fuel; street lighting; industrial plasma torches and etching processes; and lightning discharges. Needless to say, plasmas play a central role in the present school, providing the medium to support very large, travelling-wave field structures for the purposes of accelerating particles to high energies. Table \ref{lpps} provides a brief overview of these various plasma types and their properties.

\begin{table}[h]
\begin{center}
\caption{Densities and temperatures of various plasma types.}
\begin{tabular}{lcc}
\hline\hline
  \bfseries  Type \hspace{3cm}   & \bfseries Electron density   & \bfseries Temperature  \\
           							  &  \pmb{$n_e$} \bfseries (\cmcub)         &  \pmb{$T_e$} \bfseries (eV$^*$) \\ \hline 
Stars    &  $10^{26}$      &  $2\times 10^{3}$ \\
Laser fusion    &  $10^{25}$      &  $3\times 10^{3}$ \\
Magnetic fusion     &  $10^{15}$     & $10^{3}$ \\
Laser-produced  &  $10^{18} -10^{24}  $      &  $10^{2}-10^{3}$\\
Discharges  &  $ 10^{12}$       &  1-10 \\
Ionosphere & $ 10^{6}$       &  0.1 \\
ISM &  1       &  $10^{-2}$ \\
\hline \hline
$^*$ 1eV $\equiv$ 11600K
\end{tabular}
\end{center}
\label{lpps}

\end{table}

\subsection{Debye shielding} \label{Debye_sheath}

\begin{figure}[ht]
\begin{center}
\includegraphics[totalheight=1.6in]{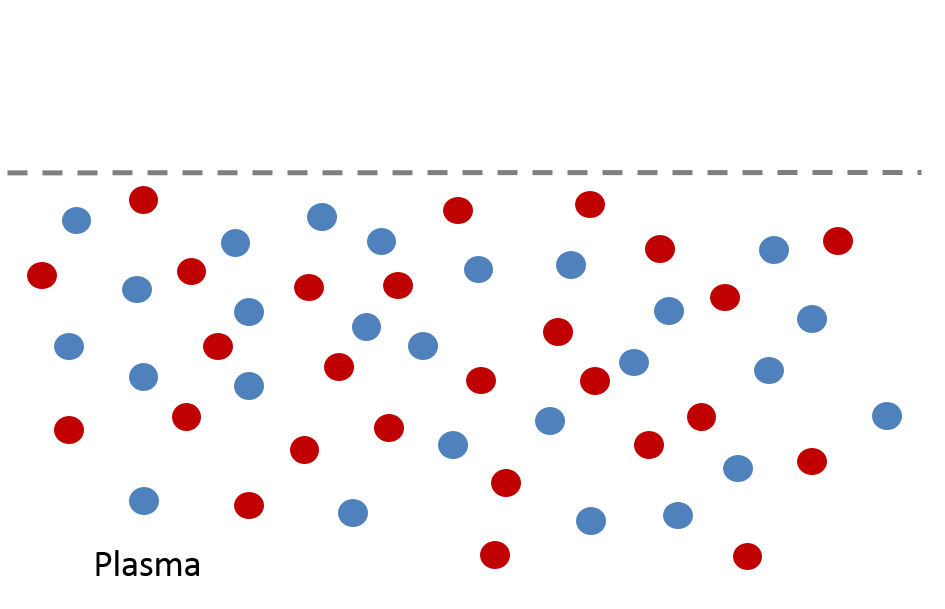}
 \hspace{0.5cm}\includegraphics[totalheight=1.5in]{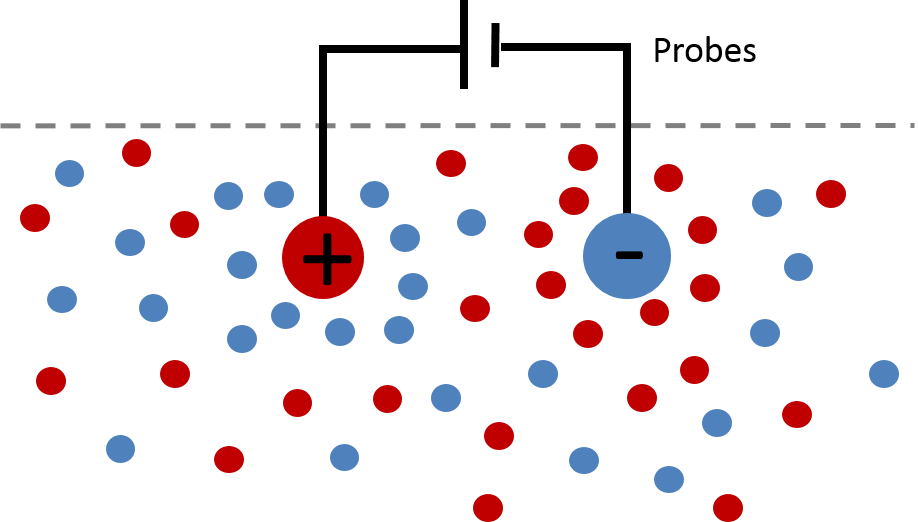}\\
 \caption{Debye shielding of charged spheres immersed in a plasma}
 \label{probes}
\end{center}
\end{figure}

In most types of plasma, quasi-neutrality is not just an ideal equilibrium state, it is something that
the~plasma actively tries to achieve by readjusting the local charge distribution in response to a disturbance. Consider a hypothetical experiment in which an ion or positively charged ball is immersed into a plasma -- see Fig.~\ref{probes}.  After some time, the ions in the ball's vicinity will be repelled and the electrons attracted, leading to an altered average charge density in this region.  In turns out that we can calculate the potential $\phi(r)$ of this sphere after such a readjustment has taken place.

First of all, we need to know how fast the electrons and ions actually move. For equal electron and ion temperatures ($T_e=T_i$), we have by definition:
\be
\fhalf m_e \overline{v}_{e}^2 = \fhalf m_i \overline{v}_{i}^2 = \frac{3}{2} k_BT_e,
\ee
where $\overline{v}_{e}$ and $\overline{v}_{i}$ represent the respective average election and ion velocities. Therefore, for a hydrogen plasma, where the ion charge and atomic number are both unity, $Z=A=1$, we find:
$$
\frac{\overline{v}_i}{\overline{v}_e}=\left(\frac{m_e}{m_i}\right)^{\half} = \left(\frac{m_e}{Am_p}\right)^{\half} \simeq \frac{1}{43}.
$$
In other words, ions are \textit{almost stationary} on the electron timescale. To a good approximation, we can often write:
\be
n_i \simeq n_0,
\label{ions_n0}
\ee
where the material (e.g. gas) number density,  $n_0=N_A\rho_m/A$ and $\rho_m$ is the usual mass density; $N_A$ the~Avogadro number.  In thermal equilibrium, the electron density follows a Boltzmann distribution \cite{chen:book}:
\be
n_e = n_i\exp(e\phi/k_BT_e),
\label{Boltzmann}
\ee
where $n_i$ is the ion density, $k_B$ is the Boltzmann constant, and $\phi(r)$ is the potential created by the external disturbance.  From Gauss' law (Poisson's equation), we can also write down:
\be
\nabla^2\phi = -\frac{\rho}{\eps0}=-\frac{e}{\eps0}(n_i-n_e).
\label{Poisson}
\ee
So now we can combine Eq. (\ref{Poisson}) with Eqs. (\ref{Boltzmann}) and (\ref{ions_n0}) in spherical geometry\footnote{$\nabla^2\rightarrow\frac{1}{r^2}\frac{d}{dr}({\tiny r}^2\frac{d\phi}{dr})$} to eliminate $n_e$ and arrive at a physically meaningful solution:
\be
\phi_D = \frac{1}{4\pi\eps0 }\frac{e^{-r/\lambda_D}}{r}.
\label{pot-debye}
\ee
This latter condition supposes that $\phi\rightarrow 0$ at $r=\infty$.  The characteristic length scale $\lambda_D$ inside the~exponential factor is known as the \textit{Debye length}, given by:
\be
\lambda_D = \left(\frac{\eps0k_BT_e}{e^2n_e}\right)^\half 
= 743 \left(\frac{T_e}{\mbox{eV}}\right)^\half\left(\frac{n_e}{\mbox{\cmcub}}\right)^{-\half} \mbox{cm}. 
\label{lambdaD}
\ee
The Debye length is a fundamental property of nearly all plasmas of interest and depends equally on its temperature and density.  An \textit{ideal} plasma has many particles per Debye sphere:
\be
N_D \equiv n_e\frac{4\pi}{3}\lambda_D^3 \gg 1,
\label{N_D}
\ee
a prerequisite for the collective behaviour encountered earlier. An alternative way of expressing this is via the so-called \textit{plasma parameter}:
\be
g \equiv \frac{1}{n_e\lambda_D^3},
\label{pl_parameter}
\ee
which is basically the reciprocal of $N_D$. Classical plasma theory is based on assumption that $g\ll 1$, which also implies dominance of collective effects over collisions between particles. Before we return to refine our plasma classification therefore, it is worth having a quick look at the nature of collisions between plasma particles.

\subsection{Collisions in plasmas}
Where $N_D \leq 1$, screening effects are reduced and collisions will dominate the particle dynamics.  In intermediate regimes, collisionality is usually measured via the \textit{electron-ion collision rate}, given by:
\be
\nu_{ei} = \displaystyle \frac{\pi^\frac{3}{2}n_e Ze^4\ln\Lambda}{2^\frac{1}{2}(4\pi\varepsilon_0)^2m_e^2v_{te}^3} \mbox{s}^{-1},
\label{nu_ei}
\ee
where $v_{te}\equiv \sqrt{k_BT_e/m_e}$ is the electron thermal velocity and $\ln\Lambda$ is a slowly varying term, the Coulomb logarithm, which typically takes on a numerical value $O(10-20)$.  The numerical coefficient in expression (\ref{nu_ei}) may vary in textbooks depending on the definition taken. This one is consistent with  Refs. \cite{kruer:book} and \cite{huba:book}, which define the collision rate according to the average time taken for a thermal electron to be deflected by $90^o$ via multiple scatterings from fixed ions. The collision frequency can also be written as 
$$\frac{\nu_{ei}}{\omega_p} \simeq \frac{Z\ln\Lambda}{10 N_D}; \;\; \mbox{with} \;\;\ln\Lambda\simeq 9N_D/Z,$$
where $\omega_p$ is the electron plasma frequency to be defined shortly in Eq. (\ref{omega-p}).

\subsection{Plasma classification}
\begin{figure}[ht]
\begin{center}
\includegraphics[totalheight=3.0in]{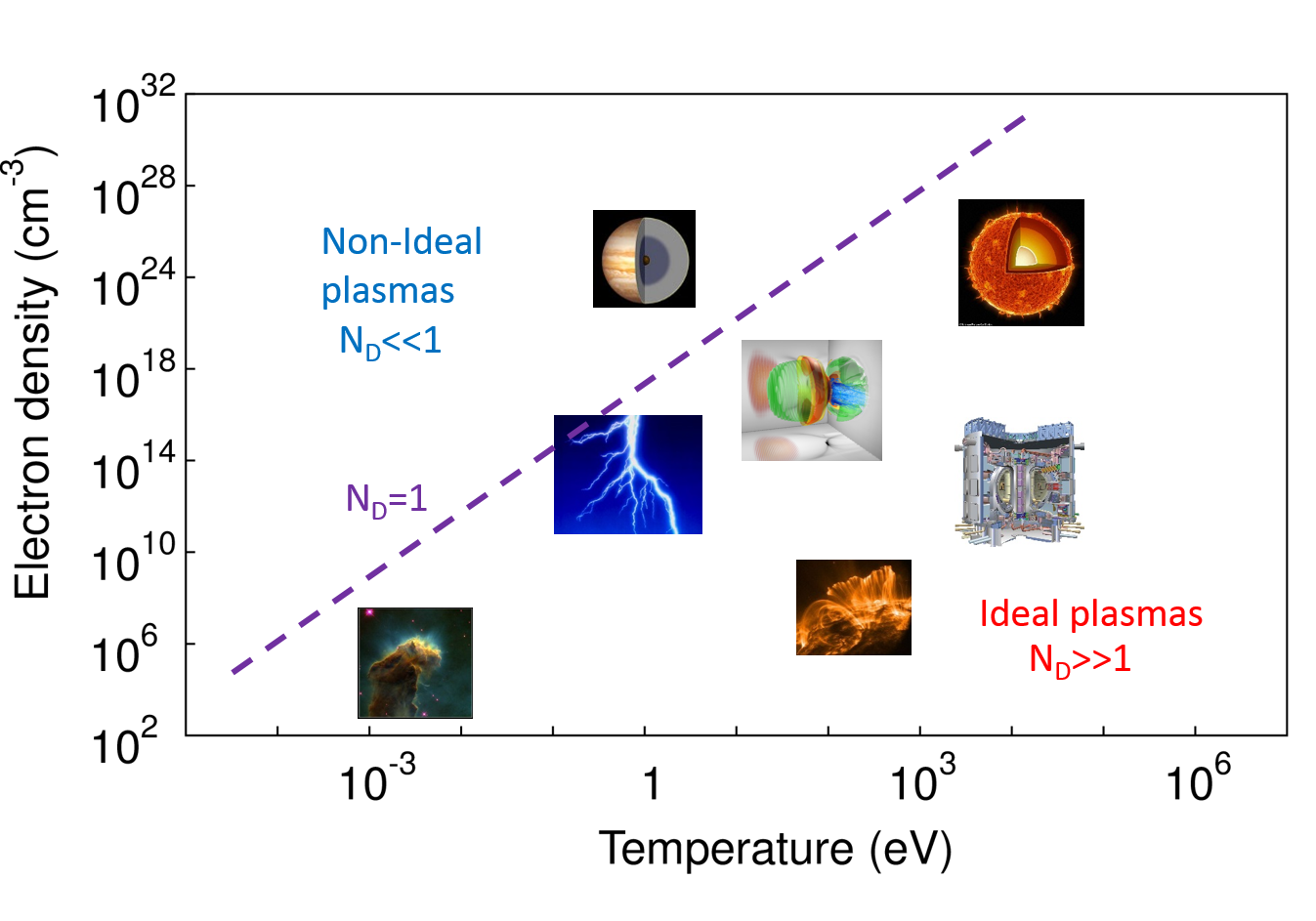}\\
\caption{Examples of plasma types in the density-temperature plane}
\label{den-temp2}
\end{center}
\end{figure}

Armed with our definition of plasma ideality (Eq. (\ref{N_D})), we can proceed to make a classification of plasma types in density-temperature space. This is illustrated for a few examples in Fig.~\ref{den-temp2}: the `accelerator' plasmas of interest to the present school are found right in the middle of this chart, having densities corresponding to roughly atmospheric pressure and temperatures of a few eV ($10^4$ Kelvin) as a result of field ionization -- see Section \ref{field_ioniz}.

\subsection{Plasma oscillations} 
\begin{figure}[ht]
\begin{center}
\includegraphics[totalheight=1.7in]{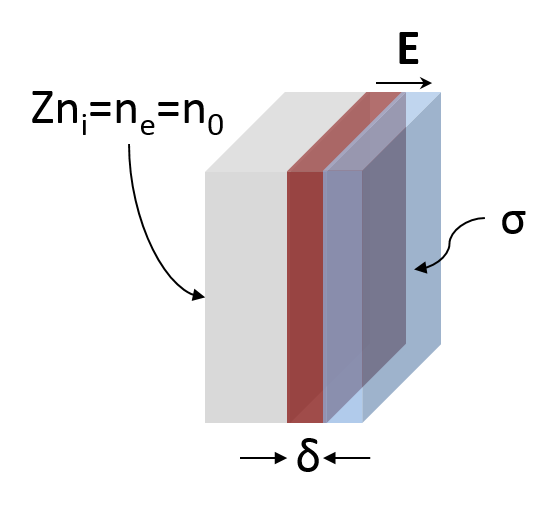}\\
\label{Langmuir-osc}
\caption{Slab or capacitor model of oscillating electron layer.}
\label{slab}
\end{center}
\end{figure}

So far we have considered characteristics, density and temperature, of a plasma in equilibrium. We can also ask how fast the plasma responds to some external disturbance, which can be due to electromagnetic waves (eg laser pulse), or particle beams.  Consider a quasi-neutral plasma slab in which an electron layer is displaced from its initial position by a distance $\delta $ -- Fig. \ref{slab}. 
This creates two 'capacitor' plates with surface charge $\sigma = \pm en_e\delta $, resulting in an electric
field:
$$
\bm{E} = \frac{\sigma}{\eps0} = \frac{en_e\delta }{\eps0}.
$$
The electron layer is accelerated back towards the slab by this restoring force according to:
$$
m_e \frac{dv}{dt} = -m_e \frac{d^2\delta }{dt^2} = -eE = \frac{e^2n_e\delta }{\eps0}.
$$
Or: $$\frac{d^2\delta }{dt^2} + \omega_p^2\delta  = 0,$$ where
\be
\omega_p \equiv \left(\frac{e^2n_e}{\varepsilon_0m_e}\right)^{\half} \simeq 5.6\times 10^4\left(\frac{n_e}{\mbox{\cmcub}}\right)^\half \mbox{s}^{-1}.
\label{omega-p}
\ee
is the \textit{electron plasma frequency}.

This quantity can be obtained via another route by returning to the Debye sheath problem of Section \ref{Debye_sheath} and asking how quickly it takes the plasma to adjust to the insertion of the foreign charge.
 \begin{figure}[ht]
\begin{center}
\includegraphics[totalheight=1.7in]{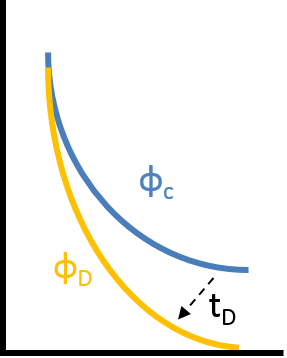}\hspace{1cm}\includegraphics[totalheight=1.6in]{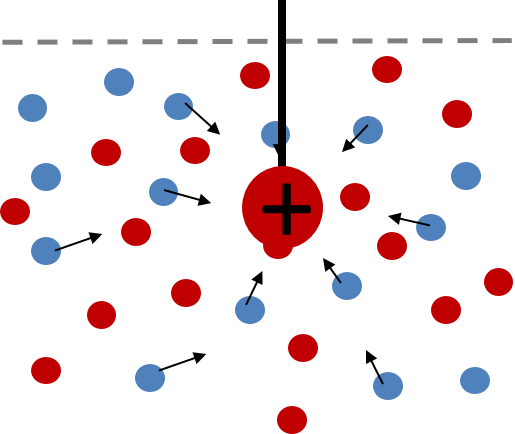}
\end{center}
\caption{Response time to form a Debye sheath}
\end{figure}
For a plasma with temperature $T_e$,  the reponse time to recover quasi-neutrality is just the ratio of \newline the Debye length to the  thermal velocity $v_{te}\equiv \sqrt{k_BT_e/m_e}$:
$$
t_D \simeq \frac{\lambda_D}{v_{te}} =  \left(\frac{\varepsilon_0k_BT_e}{e^2n_e}\cdot\frac{m}{k_BT_e}\right)^\half
= \omega_p^{-1}.
$$

If the plasma response time is shorter than the period of a external electromagnetic field (such as a laser), then this radiation will be \textit{shielded out}.  To make this more quantitative, consider ratio:
$$
\frac{\omega_p^2}{\omega^2} = \frac{e^2n_e}{\varepsilon_0m_e}\cdot \frac{\lambda^2}{4\pi^2c^2}.
$$
Setting this to unity defines the wavelength for which $n_e=n_c$, or
\be 
n_c \simeq 10^{21}\lambda_\mu^{-2}\mbox{\cmcub},
\label{n-c}
\ee
above which radiation with wavelengths $\lambda>\lambda_\mu$ will be reflected.  In the pre-satellite/cable era of the~20th century, this property was exploited to good effect in the transmission of long-wave radio signals, which utilises reflection from ionosphere to extend its reception range.
\begin{figure}[h]
\begin{center}
\includegraphics[totalheight=1.7in]{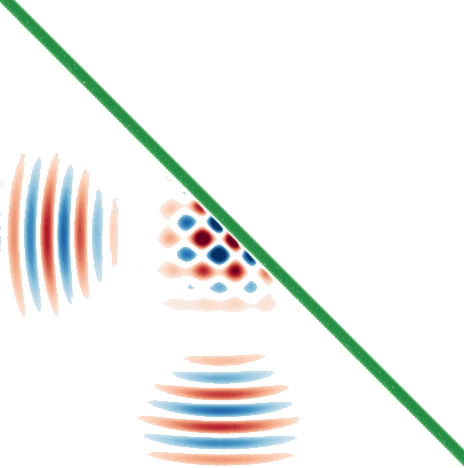} \hspace{2cm} \includegraphics[totalheight=1.7in]{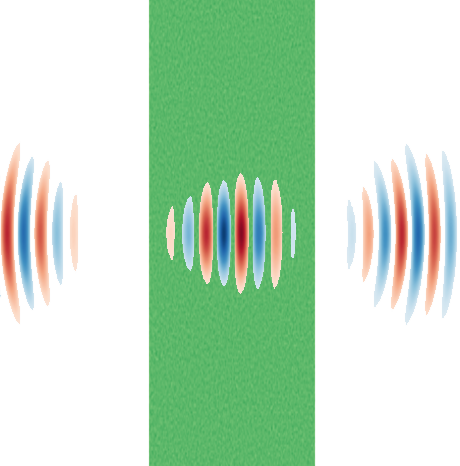}\\
\caption{Left: overdense plasma, $\omega<\omega_p$, showing mirror-like behaviour; right: underdense case, $\omega>\omega_p$, where plasma behaves like a nonlinear refractive medium. In both cases the incoming laser pulse is focussed from the~left onto the target.} 
\label{odense_udense}
\end{center}
\end{figure}

Typical gas jets have $P\sim 1$~bar; $n_e = 10^{18}-10^{19}$\cmcub\, and from Eq. (\ref{odense_udense}), the critical density for a glass laser is $n_c(1 \mu) = 10^{21}$\cmcub.  Gas-jet plasmas are therefore \textit{underdense}, since $\omega^2/\omega_p^2 = n_e/n_c \ll 1$.  In this context, \textit{collective effects} are important if $\omega_p \tau\downbox{int} > 1$, where $\tau\downbox{int}$ is some characteristic interaction time - for example the duration of a laser pulse or particle beam entering the plasma. For example, if $\tau\downbox{int} = 100$~fs, and $n_e = 10^{17}$\cmcub\ then we have $\rightarrow \omega_p \tau\downbox{int}= 1.8$, and we will need to consider the plasma response on the interaction timescale.  Generally this is the situation we seek to exploit in all kinds of plasma applications: short-wavelength radiation;  nonlinear refractive properties; generating high electric or magnetic fields; and of course, for particle acceleration.

\subsection{Plasma creation} \label{field_ioniz}
Plasmas are created via ionization, which can occur in a number of ways: through collisions of fast particles with atoms; photoionization via electromagnetic radiation, or via electrical breakdown in strong electric fields.
The latter two are examples of \textit{field ionization}, which is the mechanism most relevant to the plasma accelerator context. To get some idea of when this occurs, we need to know the typical field strength required to strip electrons away from an atom.  At the Bohr radius\index{Bohr~radius} 
\bes
a_B = \frac{4\pi\varepsilon_0\hbar^2}{me^2} = 5.3\times 10^{-9} \,\, \mbox{cm}, 
\ees
the electric field strength is:
\bea
E_a    &=& \frac{e}{4\pi\varepsilon_0a_B^2} \hspace{1.6cm}  \nonumber \\
    &\simeq& 5.1\times 10^9 \,\, \mbox{Vm}^{-1}. 
\eea
This threshold can be expressed as the so-called \textit{atomic intensity}:
\bea
I_a     &=& \displaystyle\frac{\varepsilon_0 c E_a^2 }{2}\hspace{1.6cm}  \nonumber \\
    \label{I-atomic}
    &\simeq& \displaystyle 3.51\times 10^{16} \,\, \mbox{\Wcm}. 
\eea
A laser intensity of $I_L > I_a$ will therefore guarantee ionization for any target
material, though in fact this can occur well below this threshold value (eg: $\sim 10^{14}$ \Wcm\ for hydrogen) via \textit{multiphoton} effects.  Simultaneous field ionization of many atoms produces a plasma with electron density $n_e$, temperature $T_e \sim 1-10$~eV.

\subsection{Relativistic threshold}

Before we tackle the topic of wave propagation in plasmas, it is useful to have some idea of the strength of the external fields used to excite them. To do this we resort to the classical equation of motion for an~electron exposed to a linearly polarized laser field $\bm{E} =\hat{y}E_0\sin\omega t$:
\bea
\ddt{v} &\simeq& \frac{-eE_0}{m_e} \sin\omega t, \nonumber
\eea
which implies that the electron will acquire a velocity 
\bea
\rightarrow v &=& \frac{eE_0}{m_e\omega} \cos\omega t = v\downbox{os} \cos\omega t.
\eea
This is usually expressed in terms of a dimensionless oscillation amplitude:
\be
\large
a_0 \equiv \frac{v\downbox{os}}{c} \equiv \frac{p\downbox{os}}{m_ec} \equiv \frac{eE_0}{m_e\omega c}.
\label{vos}
\ee
In articles and books this is often referred to as the `quiver' velocity or momentum and can exceed unity. In this case, normalised momentum (3rd term) is more appropriate, since the real particle velocity is just pinned to the speed of light.
The laser intensity $I_L$ and wavelength $\lambda_L$ are related to $E_0$ and $\omega$ by:
$$
I_L = \fhalf \varepsilon_0cE_0^2; \;\;  \lambda_L = \frac{2\pi c}{\omega}.
$$
Substituting these into Eq. (\ref{vos}) one obtains:
\bea
I_L &=& \frac{2\pi^2\eps0 m^2c^5}{e^2}\frac{a_0^{\; 2}}{\lambda_L^2} \nonumber \\
&\simeq & 1.37\times 10^{18} a_0^{\; 2} \lambda_\mu^2 \mbox{\Wcm}
\label{Irel}
\eea
 Conversely,
\be
a_0 \simeq 0.85 (I\downbox{18}\lambda_\mu^2)^\half,
\label{a0}
\ee
where 
$$
I\downbox{18}=\frac{I_L}{10^{18}\mbox{\Wcm}}; \;\; \lambda_\mu = \frac{\lambda_L}{\mu m}.
$$
From this expression we see that we will already have relativistic electron velocities,
 or $v\downbox{os} \sim c$, for intensities $I_L\geq 10^{18}$\Wcm, at wavelengths $\lambda_L \simeq 1$\mum.  Comparing this to \textit{thermal} velocities $v_{te}/c = \sqrt{k_BT_e/m_ec^2} = 0.01$ for $T_e=50$eV,  we see that at relativistic laser intensities, the laser field will normally completely dominate the electron motion over the ambient plasma temperature.

\section{Electron dynamics in electromagnetic waves}
As we have already seen, real plasma dynamics involves a collective response of the constituent particles to the influence of external fields. However, it is still helpful to first examine how single electrons respond to electromagnetic laser fields before tackling the more complex `many particle' problem.
\subsection{Motion in EM plane wave}
 \begin{figure}[ht]
\begin{center}
	\includegraphics[totalheight=0.8in]{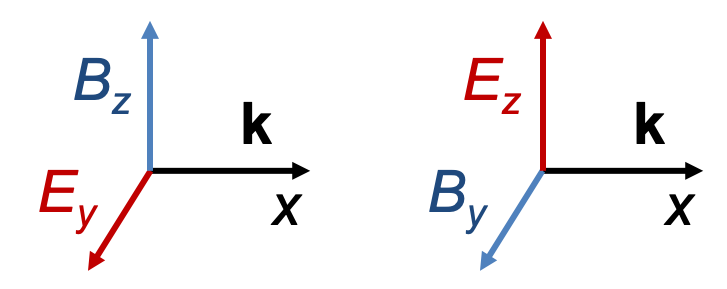}
	\caption{Geometry for treating motion in one-dimensional plane wave}
\label{wave_geom_EP}
\end{center}

\end{figure}
The simplest model of a laser field starts with a plane-wave geometry as in Fig.~\ref{wave_geom_EP}, with the transverse electromagnetic fields $\bm{E}_L=(0,E_y,E_z); \bm{B}_L=(0,B_y,B_z)$. These wave can be described equivalently by a general, elliptically polarized vector potential $\bm{A}(\omega,\bm{k})$ travelling in the positive $x$-direction :
\be
\bm{A} = A_0(0, \delta  \cos\phi, (1-\delta^2)^{\frac{1}{2}}\sin\phi),
\label{e_pol}
\ee
where $\phi = \omega t - kx$ is the phase of the wave; $A_0$ its
amplitude ($v_{os}/c=eA_0/mc$) and $\delta$ the polarization parameter. For linear polarization (LP),  $\delta = \pm 1, 0 $, we have 
$$\bm{A} = \pm\hat{\bm{y}}A_0 \cos\phi; \; \bm{A} = \hat{\bm{z}}A_0\sin\phi,$$
whereas for circularly polarized (CP) light,  $ \delta = \pm \frac{1}{\sqrt{2}}$, the vector potential becomes:
$$\bm{A} = \frac{A_0}{\sqrt{2}}(\pm\hat{\bm{y}} \cos\phi+\hat{\bm{z}}\sin\phi).$$

\noindent The electron momentum in electromagnetic wave with fields $\bm{E}$
and $\bm{B}$ given by Lorentz equation:
\be
\frac{d\bm{p}}{dt} = -e(\bm{E} + \bm{v \times B}),
\label{lorentz1}
\ee
with $\bm{p}=\gamma m\bm{v}$, and relativistic factor $\gamma =(1 + p^2/m^2c^2)^\frac{1}{2}$. This has an associated energy equation, after taking dot product of $\bm{v}$ with \eq{lorentz1}:
\be
\ddt{ }\left(\gamma mc^2\right) = -e(\bm{v\cdot E}). 
\label{energy}
\ee

\noindent The solution to this problem is treated in many texts \cite{bardsley:pra89,hartemann:pre95,gibbon:book}, so we give just a simple recipe here:

\begin{enumerate}
	\item Compute laser fields from vector potential with given polarization: $\bm{E} = -\partial_t\bm{A}, \;\bm{B}= \bm{\nabla} \times \bm{A}$ 
	\item Use dimensionless variables such that  $\omega = k = c = e = m = 1$  \\(eg: $\bm{p} \rightarrow \bm{p}/mc, \; \bm{E} \rightarrow e\bm{E}/m\omega c $, etc.) 
	\item Evaluate first integrals to yield conservation relations: $\bm{p}_\perp = \bm{A},\;\;  \gamma - p_x = \alpha$,
	where $\gamma^2 - p_x^2 - p_\perp^2 = 1$; $\alpha=$ const. 
	\item Change of variable to wave phase $\phi=t-x$ 
	\item Solve for $\bm{p}(\phi)$ and $\bm{r}(\phi)$.
\end{enumerate}

\begin{figure}[ht]
	\begin{center}
		\includegraphics[totalheight=2.5in]{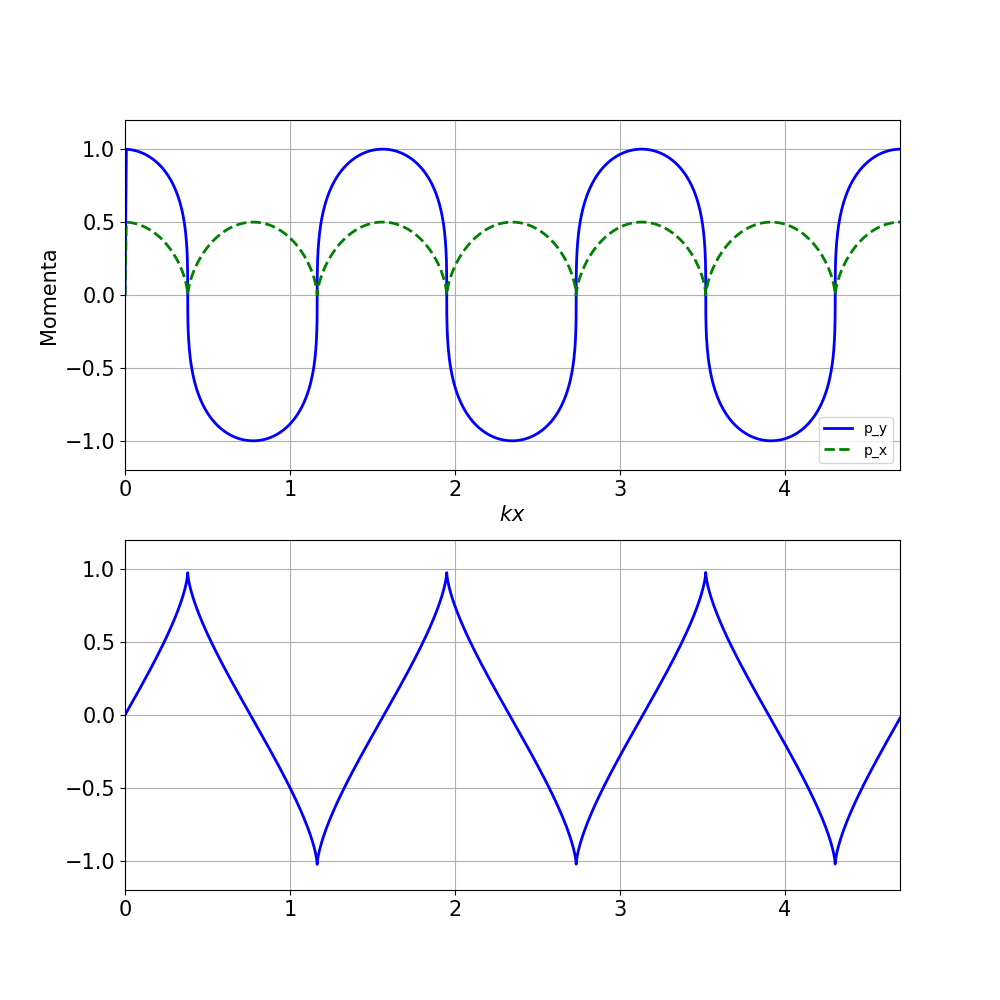}\includegraphics[totalheight=2.5in]{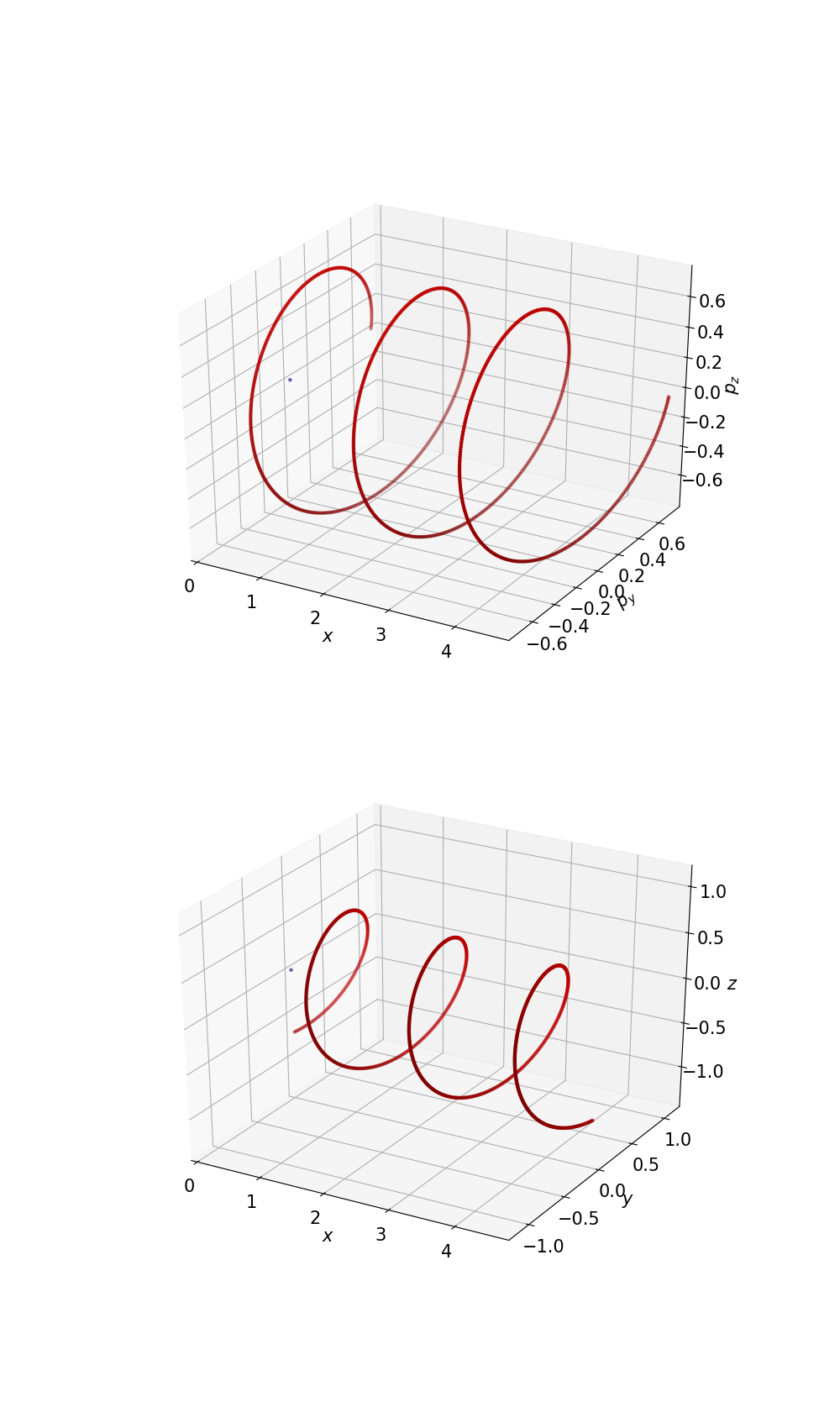}
	\end{center}
	\caption{Electron motion in LP plane wave ($\delta=1$, left); CP plane wave ($\delta=\pm 1/\sqrt{2}$, right)}
	\label{electron_orbits_lab}
\end{figure}	

In the normal laboratory frame, the electron is initially at rest before the EM wave arrives, so that  at $t=0$, $p_x = p_y=0$ and  $\gamma=\alpha = 1$. Then we can write
	\bea
	p_x &=& \frac{a_0^2}{4}\left[  1 + (2\delta^2-1)\cos 2\phi\right], \nonumber \\
	p_y &=& \delta a_0\cos\phi, \label{p_lab} \\
	p_z &=& (1-\delta^2)^\half a_0\sin\phi. \nonumber
	\eea
These can be integrated again to get the particle trajectories:
	\bea
	x &= &\frac{1}{4}a_0^2\left[\phi + \frac{2\delta^2-1}{2}\sin{2\phi}\right], \nonumber \\
	y & = & \delta a_0\sin\phi,\label{r_lab}\\
	z & =& -(1-\delta^2)^\half a_0\cos\phi. \nonumber
	\eea
Note that the solution is \textit{self-similar} in the variables $(x/a_0^2,y/a_0,z/a_0)$. The trajectories are illustrated graphically in Fig.~\ref{electron_orbits_lab}. In both cases the electron \textit{drifts}\index{electron~motion!drift~velocity}  with
average momentum $p_D\equiv \overline{p_x}=\frac{a_0^2}{4}$,  or velocity	$$\frac{v_D}{c} = \overline{v_x} = \frac{\overline{p_x}}{\overline{\gamma}} = \frac{a_0^2}{4+a_0^2}.$$
			
\noindent In a CP wave the oscillating $p_x$ component at $2\phi$ vanishes, but drift $p_D$ remains. The orbit is a helix with radius $kr_\perp = a_0/\sqrt{2}$, 
momentum $p_\perp/mc=a_0/\sqrt{2}$ and pitch angle $\theta_p = p_\perp/p_D = \sqrt{8}a_0^{-1}$.
		
\subsection{Finite pulse duration}
A laser pulse normally has a finite duration, but we can still apply some of the above solutions substituting a \textit{temporal envelope}   $\bm{A}(x,t) =   f(\phi)a_0\cos\phi,$ for the constant amplitude assumed before. In general though, it is straightforward (and more useful) to integrate the momentum equations numerically, as shown below in Fig.~\ref{finite_pulse}.
\begin{figure}[htb]
	\begin{center}
\includegraphics[totalheight=2.5in]{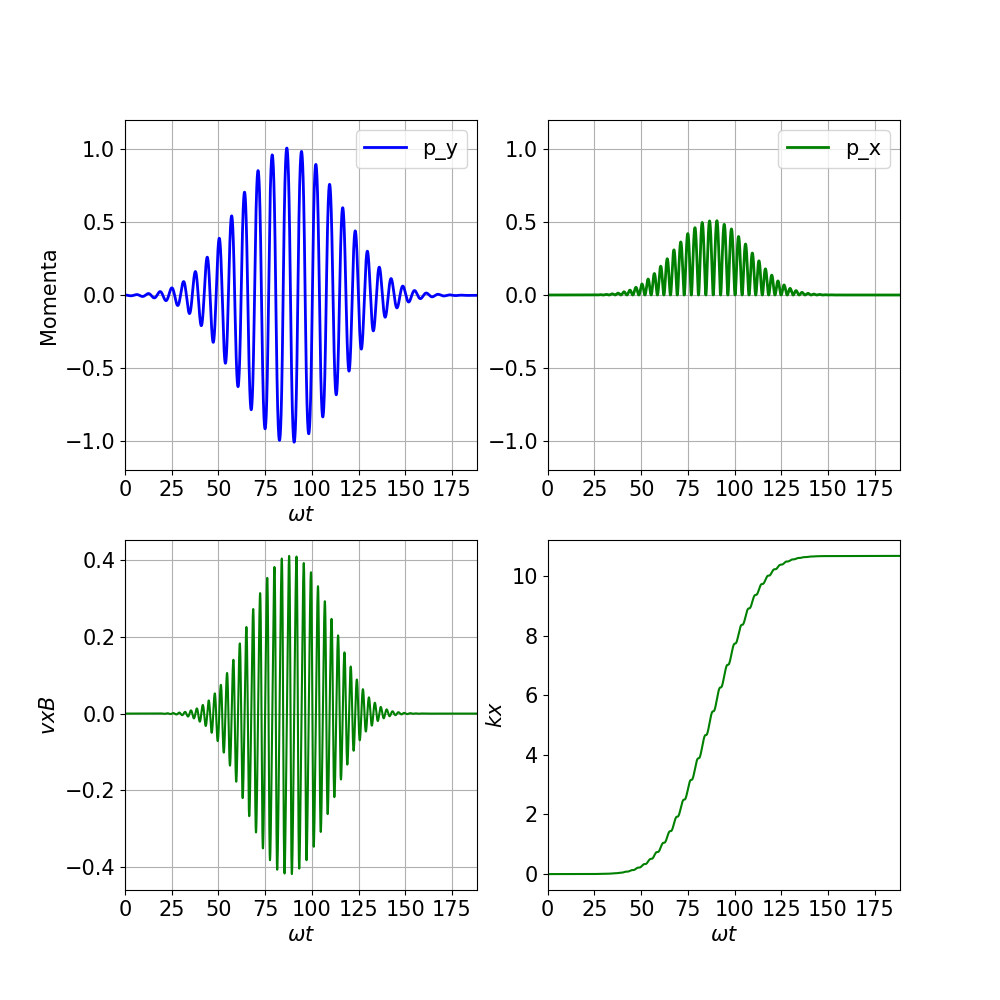}
\includegraphics[totalheight=2.5in]{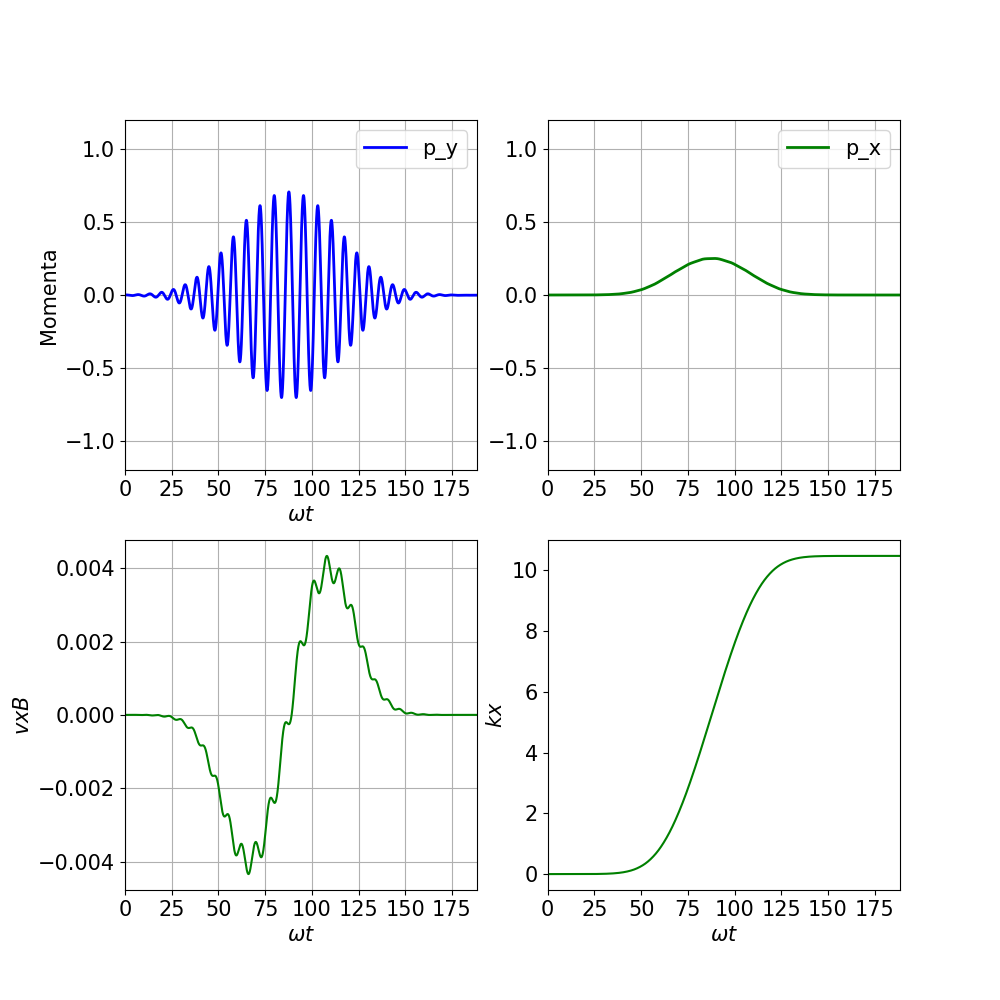}
\caption{Electron motion in laser pulse with finite number of cycles: LP pulse (left), CP pulse( right). The four insets show respectively (clockwise from top left): transverse momentum, longitudinal momentum, $v\times B$ force and longitudinal displacement.}
\label{finite_pulse}
\end{center}
\end{figure}
Note that in both cases there is no net energy gain, in agreement with the~Lawson-Woodward theorem. In the CP case the oscillations in $p_x$ are suppressed, but the drift is still there. Moreover,  the $v\times B$ oscillations also nearly vanish, but the 'DC' part, a manifestation of the \textit{longitudinal ponderomotive force}, is retained.

\subsection{Finite laser spot-size: the ponderomotive force}
The simplest way to break the symmetry of the plane wave solutions illustrated above is to introduce a~finite transverse dimension into the wave. This is the normal case for a short-pulse laser, and although we can no longer find exact solutions, we can still make a number of useful deductions about the electron motion.
Consider first a single electron oscillating slightly off-centre of focused laser beam.
		\begin{figure}[hb]
		\begin{center}
	\includegraphics[totalheight=1.7in]{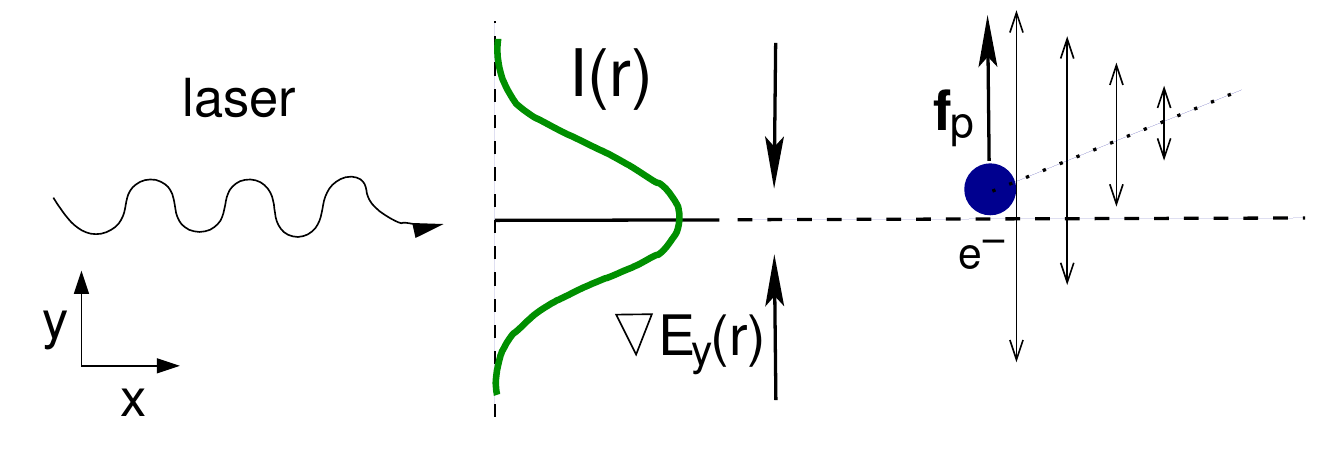}
	\caption{Schematic view of the radial ponderomotive force due to a focused beam.}
	\label{fpond_scheme}
\end{center}
		\end{figure}
 After 1st quarter-cycle, the electron sees a \emph{lower} field, and doesn't quite return to its initial position. Therefore, it is gradually accelerated away from the laser axis. Mathematically, this process can be analyzed with a~simple perturbative calculation.	In the limit $v/c\ll 1$, the equation of motion Eq. (\ref{lorentz}) for the electron becomes:
	\be
	\dbyd{v_y}{t} = - \frac{e}{m} E_y(\bm{r}).
	\label{eom_lin}
	\ee

\noindent Taylor expanding electric field about the current electron position:
	$$
	E_y(\bm{r}) \simeq E_0(y)\cos\phi + y\dbyd{E_0(y)}{y}\cos\phi + ...,
	$$
	where $\phi=\omega t - kx$ as before.\\

\noindent To lowest order, we therefore have
	$$
	v_y^{(1)} = -v\downbox{os}\sin\phi; \;\;\; y^{(1)} = \frac{v\downbox{os}}{\omega}\cos\phi,
	$$
	where $v\downbox{os} =  eE_L/m\omega $ . Substituting back into \eq{eom_lin} gives:
	$$
	\dbyd{v_y^{(2)}}{t} = - \frac{e^2}{m^2\omega^2} E_0\dbyd{E_0(y)}{y}\cos^2\phi.
	$$

\noindent Multiplying by $m$ and taking the laser cycle-average, $$\overline f=\int_0^{2\pi} f\;d\phi,$$ yields the \emph{transverse} ponderomotive force on the electron:
		\be
		f_{py} \equiv \overline{ m\dbyd{v_y^{(2)}}{t}} = -\frac{e^2}{4m\omega^2} \dbyd{E_0^2}{y}.
		\label{fpond_nonrel}
		\ee
		
\noindent A generalized relativistic version of Eq. (\ref{fpond_nonrel}) can be found by
	rewriting the Lorentz Eq. (\ref{lorentz1}) in terms of the vector potential 
	$\bm{A}$:
	\be
	\dbyd{\bm{p}}{t} + (\bm{v.}\nabla)\bm{p} = \frac{e}{c}\dbyd{\bm{A}}{t} -  
	\frac{e}{c}\bm{v}\crossb \curl \bm{A}.
	\label{lorA}
	\ee
	Make use of identity:
	\bea
	\bm{v}\crossb(\curl\bm{p}) & = & \frac{1}{m\gamma} \bm{p}\crossb\curl\bm{p} = \frac{1}{2m\gamma}\nabla\mid p\mid^2 - \frac{1}{m\gamma}(\bm{p.}\nabla)\bm{p},\nonumber 
	\eea
	separate the timescales of the electron motion into slow and fast
	components  $\bm{p}=\bm{p}^s + \bm{p}^f$ and
	average over a laser cycle, get $\bm{p}^f=\bm{A}$ and
				\be
				\bm{f}_p = \ddt{\bm{p}^s} = -mc^2\nabla\overline{\gamma},
				\label{fpond_rel}
				\ee
	where $\overline{\gamma} = \left(1+\frac{p_s^2}{m^2c^2}+ \overline{a_y^2}\right)^\half, \;\; a_y=eA_y/mc$.  
		

\section{Wave propagation in plasmas}

The theory of wave propagation is an important subject in its own right, and has inspired a vast body of literature and a number of textbooks \cite{kruer:book,dougherty:chapter,boyd:book}.  There are a great many possible ways in which plasma can support waves, depending on the local conditions, presence of external electric and magnetic fields, and so on.
Here we will concentrate on two main wave forms: longitudinal oscillations of the kind met already, and electromagnetic waves.  To derive and analyse wave phenomena, there are also a number of possible theoretical approaches depending on the length and time scales of interest, which in laboratory plasmas can range from nanometres to metres, and femtoseconds to seconds, respectively:
\begin{enumerate}
\item First principles N-body molecular dynamics
\item Phase-space methods -- Vlasov-Boltzmann equation
\item 2-fluid equations
\item Magnetohydrodynamics (single, magnetised fluid).
\end{enumerate}
The first of these approaches is rather costly and limited to much smaller regions of plasma than usually needed to describe most types of wave which supported by plasmas. Indeed, the number of particles needed for first-principles modelling of a tokamak would be around $10^{21}$; a laser-heated gas requires $10^{20}$ -- still way out of reach of even the most powerful computers available.  Clearly a more tractable model is needed and in fact, many plasma phenomena can be analysed by assuming that each charged particle component with density $n_s$ and velocity $\bm{u}_s$ behaves in a fluid-like manner, interacting with other species ($s$) via the electric and magnetic fields (method 3). The rigorous way to derive the governing equations in this approximation is via \textit{kinetic theory}, starting from method 2 \cite{kruer:book,dendy:book}, which is beyond the~scope of this lecture. Finally, slow wave phenomena on more macroscopic, ion timescales can be handled with the 4th approach above \cite{dendy:book}.  

For the present purposes we therefore begin with the 2-fluid equations for a plasma with a finite temperature ($T_e > 0$), and assumed to be collisionless ($\nu\downbox{ie}\simeq 0$) and non-relativistic, so that the fluid velocities $u \ll c$.  The equations governing the plasma dynamics under these conditions are:
\bea 
\dbyd{n_s}{t} + \nabla\cdot(n_s\bm{u}_s) &=& 0 \label{continuity}\\
n_sm_s\ddt{\bm{u}_s} &=& n_sq_s(\bm{E} + \bm{u}_s\times \bm{B}) - \nabla P_s \label{momentum-density}\\
\ddt{}(P_sn_s^{-\gamma_s}) &=& 0, \label{energy-density}
\eea
where $P_s$ is the thermal pressure of species $s$; $\gamma_s$ the specific heat ratio, or $(2+N)/N)$, where $N$ is the~number of degrees of freedom.

\noindent The continuity equation (Eq. (\ref{continuity})) tells us that (in the absence of ionization or recombination) the~number of particles \textit{of each species} is conserved.  Noting that the charge and current densities can be written $\rho_s = q_sn_s$ and  $\bm{J}_s = q_sn_s\bm{u}_s$ respectively, Eq.~(\ref{continuity}) can also be written: 
\be
\dbyd{\rho_s}{t} + \nabla\cdot\bm{J}_s = 0, \label{charge}
\ee
which expresses the conservation of \textit{charge}.

 Equation (\ref{momentum-density}) governs the motion of a fluid element of species $s$ in the presence of electric and magnetic fields $\bm{E}$ and $\bm{B}$.   In the absence of fields, and assuming strict quasineutrality ($n_e=Zn_i=n; \bm{u}_e=\bm{u}_i=\bm{u}$), we recover the more familiar \textit{Navier-Stokes} equations: 
\bea
\dbyd{\rho}{t} + \nabla\cdot(\rho\bm{u}) &=& 0, \nonumber\\
\dbyd{\bm{u}}{t} + (\bm{u}\cdot\nabla)\bm{u} &=& \frac{1}{\rho}\nabla P. \nonumber\\
\label{navier_stokes}
\eea 
By contrast, in the plasma accelerator context we will usually deal timescales on which the ions can be assumed to be motionless $\bm{u}_i=0$, and with \textit{un}magnetised plasmas, so that the momentum equation then reads:
\bea
n_em_e\ddt{\bm{u}_e} &=& -e n_e\bm{E}  - \nabla P_e
\label{electron_mom}
\eea
Note that $\bm{E}$ can include both external and internal  field components (via charge-separation).

\subsection{Longitudinal (Langmuir) waves}

\begin{figure}[hb]
\begin{center}
\includegraphics[totalheight=2.2in]{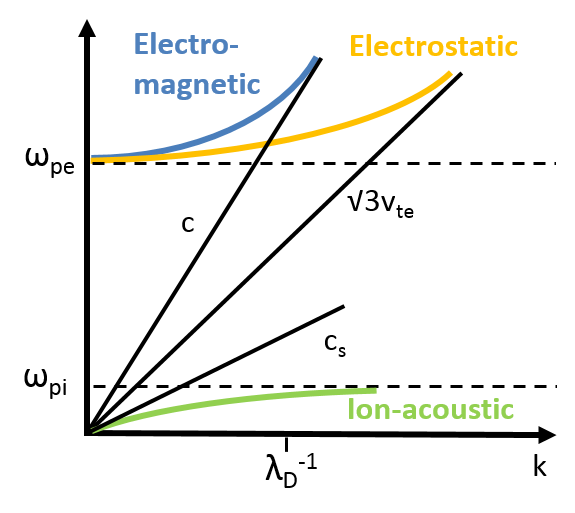}
\end{center}
\caption{Schematic illustration of dispersion relations for Langmuir, electromagnetic and ion-acoustic waves.}
\label{disp_curves}
\end{figure}

A characteristic property of plasmas is their ability to transfer momentum and energy via collective motion. One of the most important examples of this is the oscillation of the electrons against a stationary ion background, or \textit{Langmuir wave}.  Returning to the 2-fluid model, we can simplify Eqs. (\ref{continuity})--(\ref{energy-density}) by setting $\bm{u}_i=0$, restricting the electron motion to one dimension ($x$) and taking $\dbyd{}{y}=\dbyd{}{z}=0$:
\bea
\dbyd{n_e}{t}+\dbyd{}{x}(n_eu_e) &=& 0\nonumber \\
n_e\left(\dbyd{u_e}{t}+ u_e\dbyd{u_e}{x}\right) &=&  -\frac{e}{m}n_eE - \frac{1}{m}\dbyd{P_e}{x}\label{Langmuir_1d} \\
\ddt{}\left(\frac{P_e}{n_e^{\gamma_e}}\right) &=&0 \nonumber
\eea
The above system (\ref{Langmuir_1d}) has 3 equations and 4 unknowns. 
To close it we need an expression for the electric field, which, since  $\bm{B}=0$, can be found from Gauss' law (Poisson's equation) with $Zn_i=n_0$:
\be
\dbyd{E}{x} = \frac{e}{\eps0}(n_0-n_e) 
\label{Poisson_1d}
\ee
The system of equations (\ref{Langmuir_1d}) and (\ref{Poisson_1d}) is nonlinear, and  apart from a few special cases, cannot be solved exactly.  A common technique for analyzing waves in plasmas therefore is to \textit{linearize} the equations, assuming the perturbed amplitudes are small compared to the equilibrium values:
\bea
n_e &=& n_0 + n_1, \nonumber \\
u_e &=& u_1, \nonumber \\
P_e &=& P_0 + P_1, \nonumber \\
E &=& E_1, \nonumber 
\eea   
where $n_1\ll n_0, P_1\ll P_0$.  These expressions are substituted into (\ref{Langmuir_1d},\ref{Poisson_1d}) and all products $n_1\partial_t u_1, u_1\partial_xu_1$ etc. are neglected to get a set of linear equations for the perturbed quantities:
\bea
\dbyd{n_1}{t}+n_0\dbyd{u_1}{x} &=& 0,\nonumber \\
 n_0\dbyd{u_1}{t} &=& -\frac{e}{m}n_0E_1 - \frac{1}{m}\dbyd{P_1}{x}, \label{fluid_lin}\\
 \dbyd{E_1}{x} &=& -\frac{e}{\eps0}n_1, \nonumber \\
 P_1 &=& 3k_BT_en_1. \nonumber
\eea
The expression for $P_1$ results from specific heat ratio $\gamma_e=3$ and assuming isothermal background electrons, $P_0=k_BT_en_0$ (ideal gas) -- see Kruer (1988).  We can now eliminate $E_1, P_1$ and $u_1$ from Eq.~(\ref{fluid_lin}) to get:
\be
\left(\dbyd{^2}{t^2} - 3v_{te}^2\dbyd{^2}{x^2}+\omega_p^2\right)n_1 = 0,
\label{langmuir_wave}
\ee
with $v_{te}^2=k_BT_e/m_e$ and $\omega_p$ given by Eq. (\ref{omega-p}) as before.  Finally, we look for plane wave solutions of the form $A = A_0e^{i(\omega t-kx)}$, so that our derivative operators become: $\dbyd{}{t}\rightarrow i\omega; \dbyd{}{x}\rightarrow -ik$.   Substitution into Eq. (\ref{langmuir_wave}) yields the Bohm-Gross dispersion relation:
\be 
\omega^2 = \omega_p^2 + 3k^2v_{te}^2.
\label{Bohm-Gross}
\ee
This and other dispersion relations are often depicted graphically on a chart such as that in Fig.~\ref{disp_curves}, which gives an overview of which propagation modes are permitted for low- and high-wavelength limits. 

\subsection{Transverse waves}
\begin{figure}[ht]
\begin{center}
\includegraphics[totalheight=2.0in]{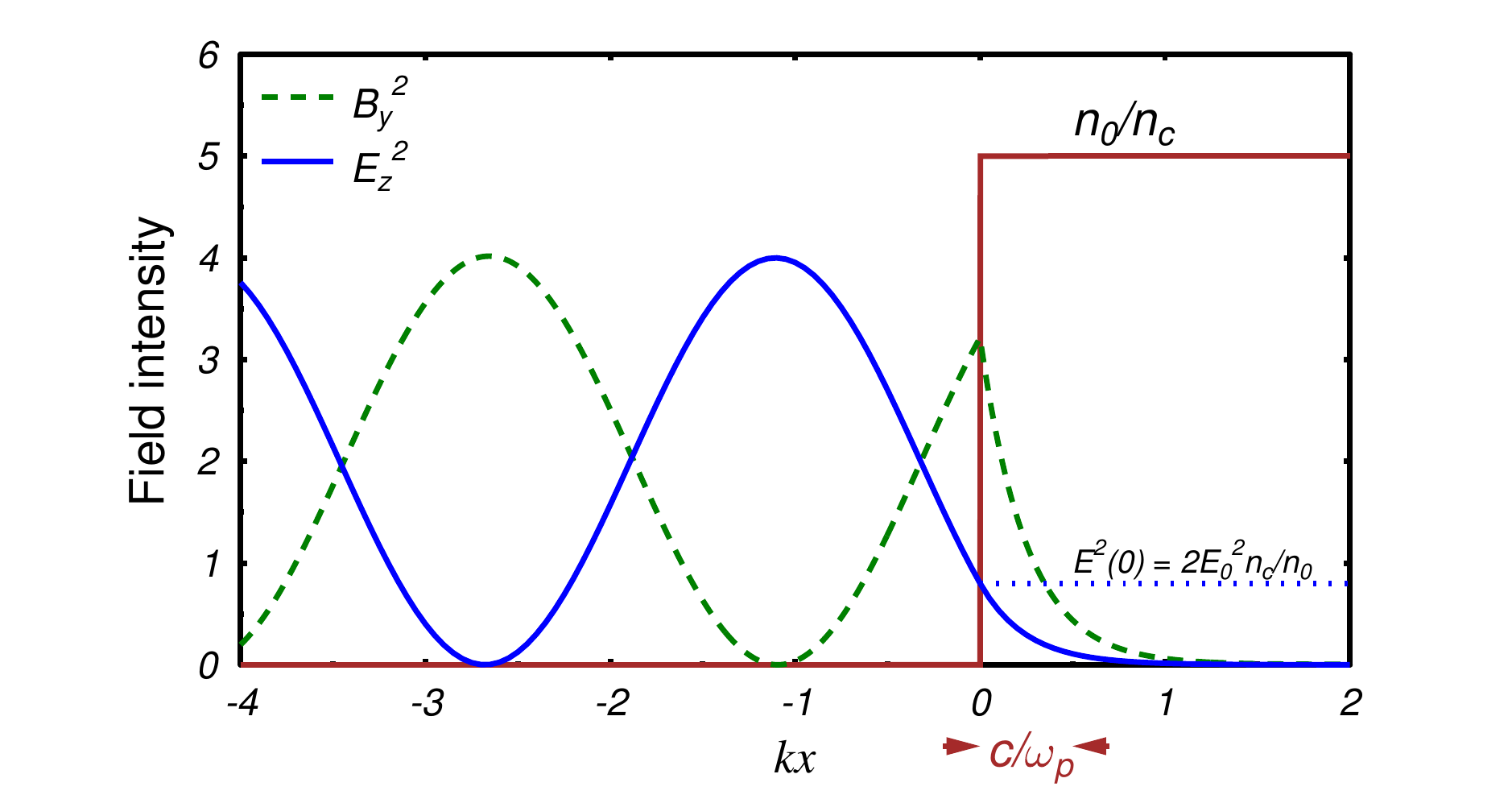}
\end{center}
\caption{Electromagnetic fields resulting from reflection of an incoming wave by an overdense plasma slab.}
\label{skin_fields}
\end{figure}

To describe \textit{transverse} electromagnetic (EM) waves, we need two additional Maxwells equations; Faraday's law (\ref{faraday}) and Amp\`ere's law (\ref{ampere}), which we will introduce properly shortly (see Eqs. (~\ref{faraday}) and (\ref{ampere})).  For the time-being it is helpful to simplify things by making use of our previous analysis of  small-amplitude, longitudinal waves. Therefore, we linearize and again apply the harmonic approximation $\dbyd{}{t}\rightarrow i\omega$:
\bea
\nabla \bm{\times E}_1 & = & - i\omega\bm{B}_1,
\label{faraday_lin}\\
\nabla \bm{\times B}_1 & = & \mu_0 \bm{J}_1 +i\eps0\mu_0\omega\bm{E}_1, 
\label{ampere_lin} 
\eea
where the transverse current density is given by:
\be
\bm{J}_1 = -n_0e\bm{u}_1.
\label{current_lin}
\ee
This time we look for pure EM plane-wave solutions with $\bm{E}_1 \perp \bm{k}$ -- see Fig.~\ref{wave_geom_EP}, and also note that the group and phase velocities are assumed to be large enough $v_p, v_g \gg v_{te}$, so that we can assume a \textit{cold} plasma with $P_e=n_0k_BT_e\simeq 0$.
%
\noindent The linearized electron fluid velocity and corresponding current are then:
\bea
\bm{u}_1 &=& -\frac{e}{i \omega m_e}\bm{E}_1, \nonumber \\
\bm{J}_1 &=& \frac{n_0e^2}{i \omega m_e} \bm{E}_1 \equiv \sigma \bm{E}_1 ,
\label{ac-conductivity}
\eea
where $\sigma$ is the \textit{AC electrical conductivity}.  By analogy with dielectric media (see e.g., Ref. \cite{jackson:book}), in which Ampere's law is usually written $\nabla \bm{\times B}_1  = \mu_0 \partial_t{\bm{D}}_1$, by substituting Eq. (\ref{ac-conductivity}) into Eq. (\ref{ampere}), can show that:
$$
\bm{D}_1 = \eps0\varepsilon\bm{E}_1,
$$
with
\be 
\varepsilon = 1 + \frac{\sigma}{i\omega\eps0} = 1-\frac{\omega_p^2}{\omega^2}.
\label{Dielectric_function}
\ee
From Eq. (\ref{Dielectric_function}) it follows immediately that:
\be 
\eta \equiv \sqrt{\varepsilon} = \frac{ck}{\omega} = \left(1-\frac{\omega_p^2}{\omega^2}\right)^{\half}, \label{refractive_index}
\ee
with
\be 
\omega^2 = \omega_p^2 + c^2k^2
\label{em_disprel}
\ee
The above expression can also be found directly by elimination of $\bm{J}_1$ and $\bm{B}_1$ from Eqs. (\ref{faraday_lin})--(\ref{ac-conductivity}).  From the dispersion relation Eq. (\ref{em_disprel}), also depicted in Fig.~\ref{disp_curves}, a number of important features of EM wave propagation in plasmas can be deduced. For \textit{underdense} plasmas ($n_e\ll n_c$):
\bea
\mbox{Phase velocity:} \hspace{.5cm} v_p &=& \frac{\omega}{k} \simeq c\left(1+\frac{\omega_p^2}{2\omega^2}\right) > c \nonumber \\ 
\mbox{Group velocity:} \hspace{.5cm} v_g &=& \dbyd{\omega}{k} \simeq c\left(1-\frac{\omega_p^2}{2\omega^2}\right) < c \nonumber 
\eea
In the opposite case, $n_e > n_c$, the refractive index $\eta$ becomes imaginary, and the wave can no longer propagate, becoming evanescent instead, with a decay length determined by the \textit{collisionless skin depth} $c/\omega_p$ -- Fig.~\ref{skin_fields}

\subsection{Nonlinear wave propagation}
\begin{figure}[htb]
			\begin{center}
				i) 
				\includegraphics[totalheight=2.5in]{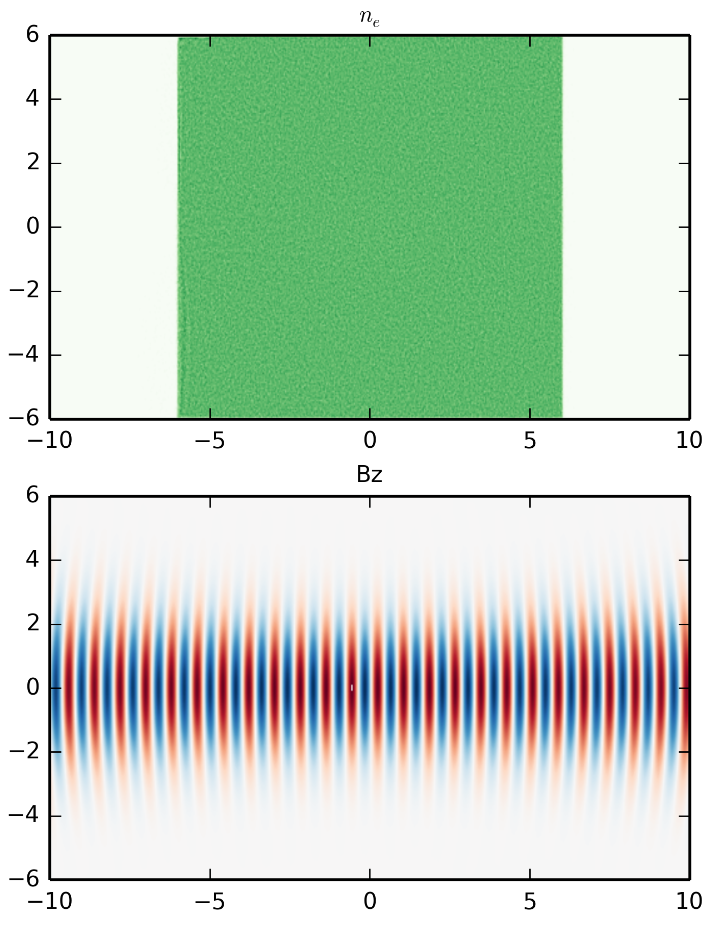}\hspace{1cm}			\includegraphics[totalheight=2.5in]{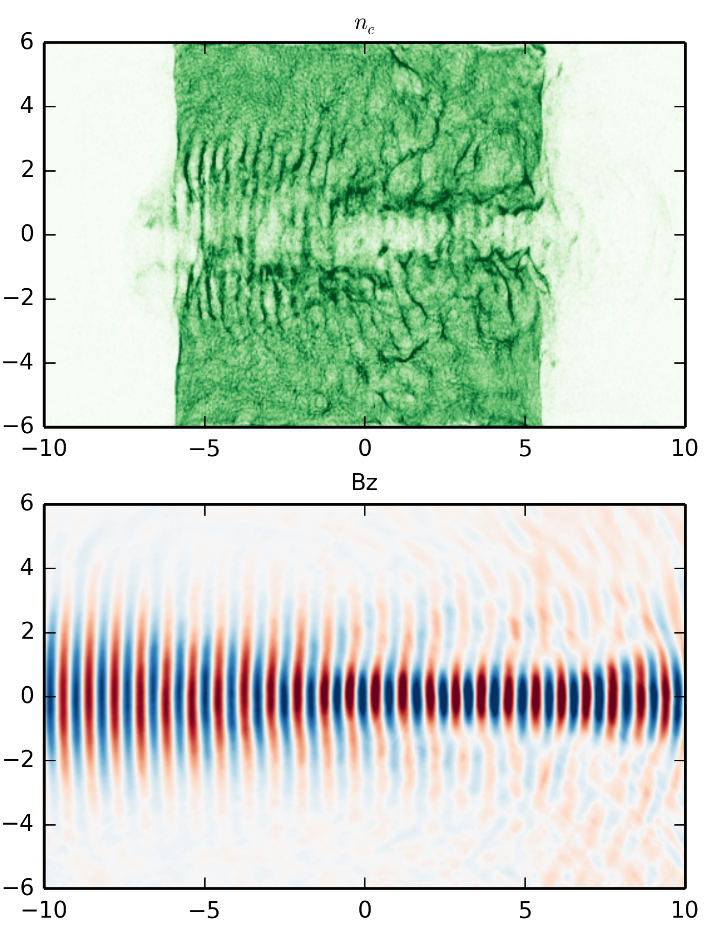}
			\end{center}
		\caption{Laser propagation in underdense plasma for  $P_L/P_c \ll 1$ (left) and $P_L=2P_c$ (right)} 
		\label{rel_propag}
\end{figure}

So far we have considered purely longitudinal or transverse waves: linearising the wave equations ensures that any nonlinearities or coupling between these two modes is excluded; a reasonable approximation for low amplitude waves, but inadequate to describe strongly driven waves in the relativistic regime of interest for plasma accelerator schemes. The starting point for most analyses of nonlinear wave propagation phenomena is the Lorentz equation of motion for the electrons in a \textit{cold} ($T_e=0$), unmagnetized plasma, together with Maxwell's equations \cite{kruer:book,gibbon:book}.  We also make two assumptions i) that the ions are initially assumed to be singly charged ($Z=1$) and are treated as a immobile ($v_i=0$), homogeneous background with $n_0=Zn_i$, and ii) that thermal motion can be neglected, since the temperature remains small compared to the typical oscillation energy in the laser field ($v\downbox{os}\gg v_{te}$).  The starting equations (SI units) are then as follows:
\bea
\frac{\partial \bm{p}}{\partial t} + (\bm{v\cdot \nabla})\bm{p} & = & -e(\bm{E} + \bm{v \times B}),       
\label{lorentz} \\
\nabla \bm{\cdot E} & = &  \frac{e}{\varepsilon_0}(n_0 - n_e),
\label{poisson}\\
\nabla \bm{\times E} & = & - \frac{\partial\bm{B}}{\partial t},
\label{faraday}\\
c^2\nabla \bm{\times B} & = & -\frac{e}{\varepsilon_0}n_e\bm{v} +  \frac{\partial\bm{E}}{\partial t}, 
\label{ampere} \\
\nabla \bm{\cdot B} & = &  0,       \label{divb}
\eea
where $\bm{p} = \gamma m_e \bm{v}$ and $\gamma = (1+p^2/m_e^2c^2)^\half$. To simplify matters we first assume a plane-wave geometry like that in Fig.~\ref{wave_geom_EP}, with the transverse electromagnetic fields given by $\bm{E}_L=(0,E_y,0); \bm{B}_L=(0,0,B_z)$.  
From Eq. (\ref{lorentz}) the transverse electron momentum is then simply given by:
\be
p_y = eA_y,
\label{canmom}
\ee
where $E_y = \partial A_y/\partial t$.  This relation expresses conservation of canonical momentum.
Substituting $\bm{E}=-\nabla\phi-\partial \bm{A}/\partial t;\; \bm{B}=\nabla\times\bm{A}$ into Amp\`ere Eq. (\ref{ampere}) yields:
$$
c^2\nabla\times (\nabla\times \bm{A}) + \frac{\partial^2 \bm{A}}{\partial t^2}  = 
\frac{\bm{J}}{\eps0} - \nabla\dbyd{\phi}{t},
$$
where the current $\bm{J} = -en_e\bm{v}$.  Now we use a bit of vectorial wizardry, splitting the current into rotational (solenoidal) and irrotational (longitudinal) parts:
$$
\bm{J} = \bm{J}_\perp + \bm{J}_{||} = \nabla\times\bm{\Pi} + \nabla{\Psi}
$$
from which we can deduce (see Jackson):
$$
\bm{J}_{||} - \frac{1}{c^2}\nabla\dbyd{\phi}{t}=0.
$$
Applying the Coulomb gauge $\nabla\cdot\bm{A}=0$ and $v_y=eA_y/\gamma$ from (\ref{canmom}), to finally get:
\be
\frac{\partial^2 A_y}{\partial t^2} - c^2\nabla^2 A_y    = \mu_0J_y = -\frac{e^2n_e}{\eps0 m_e\gamma} A_y.
\label{emwave}
\ee
The nonlinear source term on the RHS contains two important bits of physics: $ n_e = n_0 + \delta n$, which couples the EM wave to plasma waves; $\gamma=\sqrt{1+\bm{p}^2/m_e^2c^2}$ which accounts for the relativistically enhanced electron inertia.  The above wave equation thus already describes a host of effects which high-intensity laser pulses will be subjected to in a plasma:
\begin{itemize}

\item diffraction due to finite focal spot $\sigma_L$: $ Z_R=2\pi\sigma_L^2/\lambda$ 
\item ionization effects $dn_e/dt \Rightarrow$ refraction due to radial density gradients
\item relativistic self-focusing and self-modulation \\ $ \Rightarrow \eta(r) = \sqrt{\left(1-\frac{\omega_p^2(r)}{\gamma_0(r)\omega^2}\right)}$
\item ponderomotive channelling $\Rightarrow \nabla_r n_e$
\item scattering by plasma waves $\Rightarrow k_0 \rightarrow k_1 + k_p$.
\end{itemize}
All of these nonlinear effects are important for laser powers $P_L > 1 TW$. One of the most important of these is relativistic self-focussing, for which there is a power threshold \cite{litvak:1970,max:1974,sprangle:1990}. The laser power can be written:

	\bar
	P_L &=& \left(\frac{m\omega c}{e}\right)^2\left(\frac{c}{\omega_p}\right)^2\frac{c\epsilon_0}{2}\int_0^\infty 2\pi r a^2(r)dr\\
	&=& \fhalf\left(\frac{m}{e}\right)^2 c^5\epsilon_0\left(\frac{\omega}{\omega_p}\right)^2\tilde{P},  \\
	&\simeq &0.35 \left(\frac{\omega}{\omega_p}\right)^2\tilde{P}~ \mbox{GW}, \;\;\;\mbox{where} \;\;\; \tilde{P}\equiv \pi a_0^2 (\omega_p^2 \sigma_L^2/c^2).
	\ear
	The normalized critical power often quoted in early texts $\tilde{P_c}=16\pi$ thus corresponds to:
		\be
		P_c \simeq 17.5 \left(\frac{\omega}{\omega_p}\right)^2 \mbox{GW}.
		\label{Pcrit}
		\ee
For example, a laser with wavelength $\lambda_L = 0.8 \mu m$ propagating in an electron gas with $n_e = 1.6\times 10^{20}\mbox{~\cmcub}$ sees a normalized density of $\frac{n_e}{n_c} = \left(\frac{\omega_p}{\omega}\right)^2 = 0.1$ and will start to focus at the threshold power $\Rightarrow P_c = 0.175 \mbox{TW}$.
An example of this behaviour for a long pulse in an underdense plasma is shown in Fig.~\ref{rel_propag}.

Taking the \textit{longitudinal} component of the momentum Eq. (\ref{lorentz}) gives:
$$
\ddt{}(\gamma m_ev_x) = -eE_x - \frac{e^2}{2m_e\gamma}\dbyd{A_y^2}{x}.
$$
We can eliminate $v_x$ using Amp\`ere's law (\ref{ampere})$_x$:
$$
0=-\frac{e}{\eps0}n_e v_x + \dbyd{E_x}{t},
$$
while the electron density can be determined via Poisson's Eq. (\ref{poisson}):
$$
n_e = n_0 - \frac{\varepsilon_0}{e}\dbyd{E_x}{x}.
$$

The above (closed) set of equations can in principle be solved numerically for arbitrary pump strengths.  For the moment, we simplify things by linearizing the \textit{plasma} fluid quantities:
\bea
n_e &\simeq& n_0 + n_1 + ...\nonumber \\
v_x &\simeq& v_1 + v_2 + ...\nonumber 
\eea
and neglect products like $n_1v_1$ etc. This finally leads to:

\be
\left( \dbyd{^2 }{t^2} + \frac{\omega_p^2}{\gamma_0}\right) E_x = -\frac{\omega_p^2e}{2m_e\gamma_0^2}\dbyd{ }{x}A_y^2 .
\label{langmuir}
\ee
The driving term on the RHS is the \textit{relativistic ponderomotive force}, with $\gamma_0=(1+a_0^2/2)^\half$.  Some solutions of Eq.~(\ref{langmuir}) are shown in Fig.~\ref{qsa_wake} below, for low- and high-intensity laser pulses respectively. The properties of these wakes will be discussed in detail in subsequent lectures, but we can already see some obvious qualitative differences in the linear and nonlinear waveforms; the latter typically characterised by a spiked density profile, saw-tooth electric field and longer wavelength.
\begin{figure}[ht]
\begin{center}
\includegraphics[totalheight=2.in]{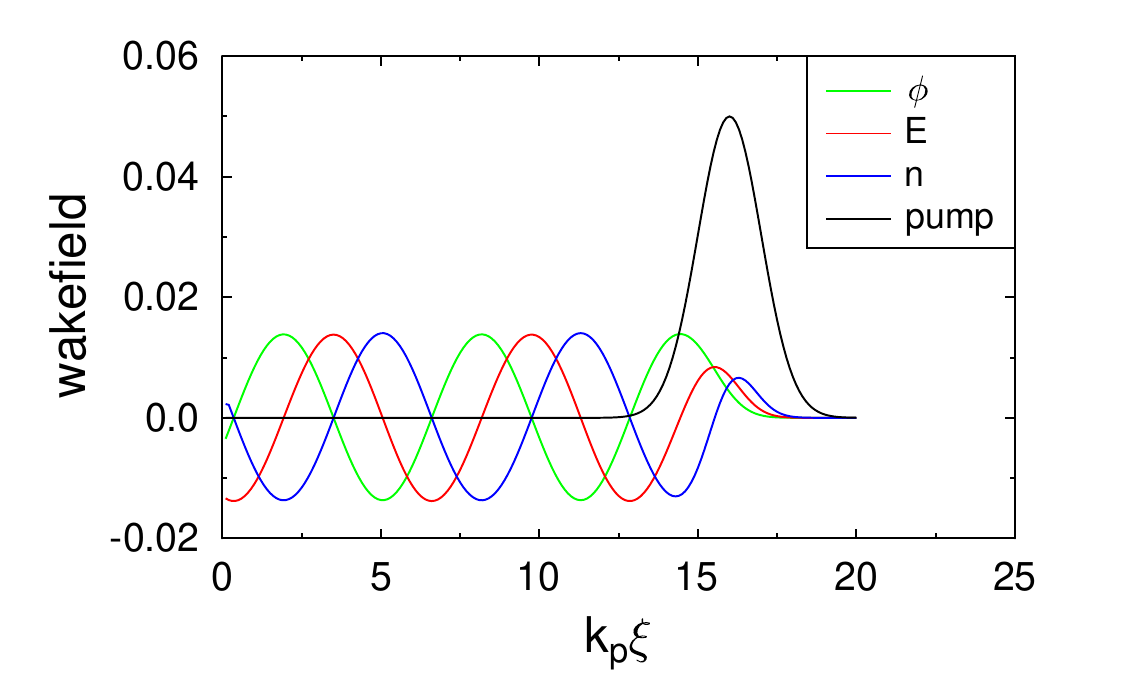}
\includegraphics[totalheight=2.in]{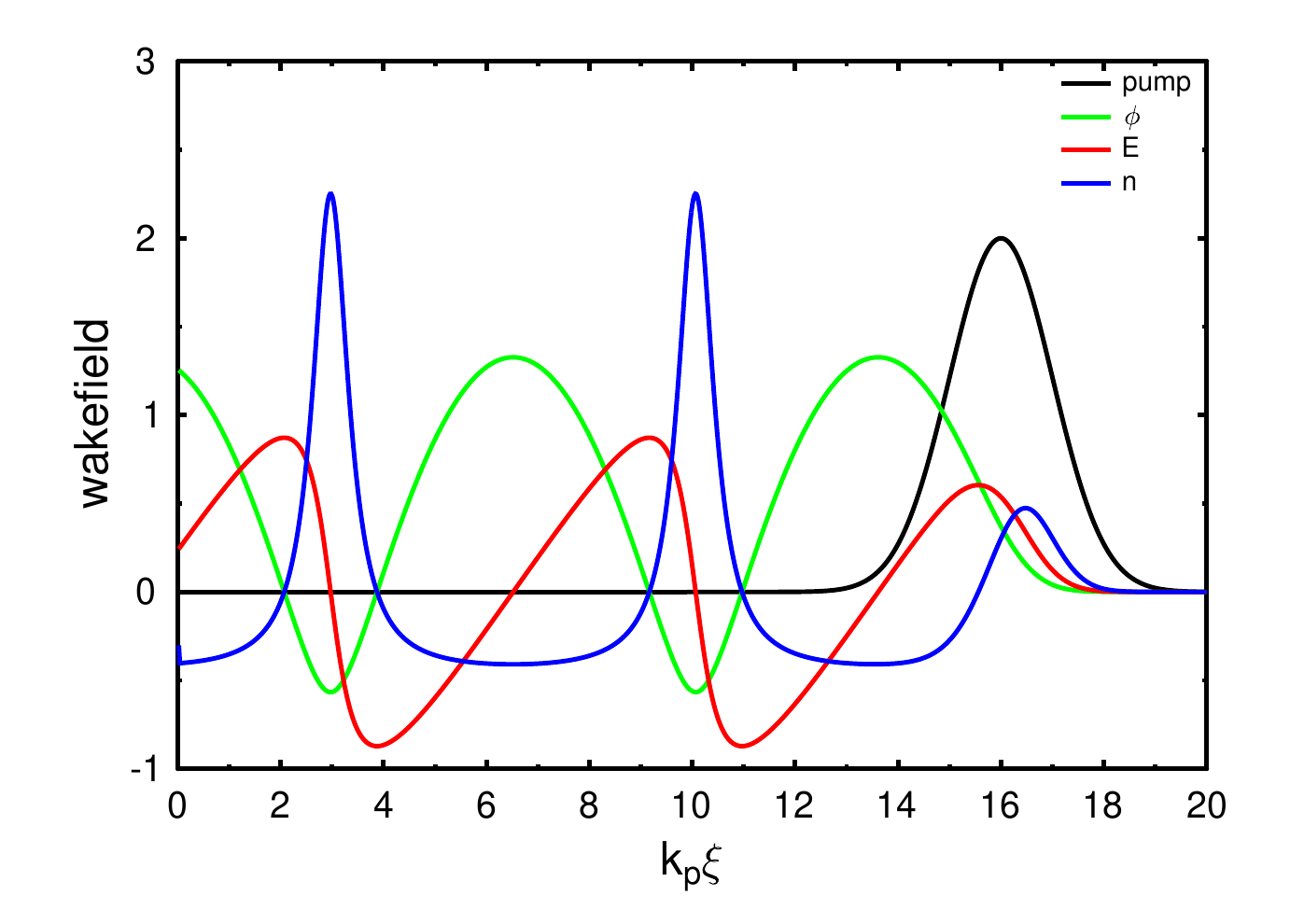}
\caption{Wakefield excitation by a short-pulse laser propagating in the positive x-direction in the linear regime (left) and nonlinear regime (right).}
\label{qsa_wake}
\end{center}
\end{figure}
The coupled fluid Eqs. (\ref{emwave}) and (\ref{langmuir}) and their fully non-linear counterparts describe a vast range of nonlinear laser-plasma interaction phenomena, many of which are treated in the later lectures of the school: plasma wake generation; blow-out regime
laser self-focussing and channelling; parametric instabilities; harmonic generation, and so on.
Plasma-accelerated particle \textit{beams}, on the other hand,  cannot be treated with fluid theory and demand a~more sophisticated kinetic approach, usually requiring the assistance of numerical models solved with the help of powerful supercomputers. 

\pagebreak

\pagebreak
\section{Useful constants and formulae}

\begin{table}[h]
\begin{center}
\caption{Commonly used physical constants}
\label{tab:constants}
\begin{tabular}{l|c|c|c}
\hline\hline
\bfseries Name & \bfseries Symbol & \bfseries Value (SI) & \bfseries Value (cgs)\\
\hline
 Boltzmann constant   &   $k_B$ & $1.38\times 10^{-23}$ JK$^{-1}$ & $1.38\times 10^{-16}$ erg K$^{-1}$  \\
Electron charge 	& $e$ 	& $1.6\times 10^{-19}$ C & $4.8\times 10^{-10}$ statcoul \\
Electron mass 	& $m_e$	& $9.1\times 10^{-31}$ kg & $9.1\times 10^{-28}$ g \\
Proton mass 		& $m_p$	& $1.67\times 10^{-27}$ kg & $1.67\times 10^{-24}$ g \\
Planck constant	& $h$	& $6.63\times 10^{-34}$ Js & $6.63\times 10^{-27}$ erg-s \\
Speed of light	& $c$	& $3\times 10^8$ ms$^{-1}$ & $3\times 10^{10}$ cms$^{-1}$\\
Dielectric constant	& $\varepsilon_0$ & $8.85\times10^{-12} $ Fm$^{-1}$ & --- \\
Permeability constant 	& $\mu_0$	& $4\pi\times10^{-7}$ & --- \\
Proton/electron mass ratio	& $m_p/m_e$ & 1836 & 1836 \\
Temperature = 1eV	& $e/k_B$  & 11604 K	& 11604 K\\
Avogadro number 		& $N_A$	   & $6.02\times 10^{23}$ mol$^{-1}$ &$6.02\times 10^{23}$ mol$^{-1}$ \\
\hline\hline
\end{tabular}
\end{center}
\end{table}

\begin{table}[h]
\begin{center}
\caption{Formulae in SI and cgs units}
\label{tab:formulae}
\begin{tabular}{l|c|c|c}
\hline\hline
\bfseries Name & \bfseries Symbol & \bfseries Formula (SI) & \bfseries Formula (cgs)\\
\hline
Debye length		& $\lambda_D$	& $\displaystyle \left(\frac{\varepsilon_0k_BT_e}{e^2n_e}\right)^\frac{1}{2}$ m
					& $\displaystyle \left(\frac{k_BT_e}{4\pi e^2n_e}\right)^\frac{1}{2}$ cm \\
Particles in Debye sphere & $N_D$ & $ \displaystyle \frac{4\pi}{3}\lambda_D^3$ & $ \displaystyle\frac{4\pi}{3}\lambda_D^3$\\
Plasma frequency (electrons)& $\omega_{pe}$  & $\displaystyle \left(\frac{e^2n_e}{\varepsilon_0 m_e}\right)^\frac{1}{2}$ s$^{-1}$	& $\displaystyle \left(\frac{4\pi e^2n_e}{m_e}\right)^\frac{1}{2}$ s$^{-1}$\\
Plasma frequency (ions)  &  $\omega_{pi}$  & $\displaystyle \left(\frac{Z^2e^2n_i}{\varepsilon_0 m_i}\right)^\frac{1}{2}$ s$^{-1}$	& $\displaystyle \left(\frac{4\pi Z^2 e^2n_i}{m_i}\right)^\frac{1}{2}$ s$^{-1}$\\
Thermal velocity & $v_{te}=\omega_{pe}\lambda_D$	& $\displaystyle \left(\frac{k_BT_e}{m_e}\right)^\frac{1}{2}$ ms$^{-1}$ & $\displaystyle \left(\frac{k_BT_e}{m_e}\right)^\frac{1}{2}$ cms$^{-1}$\\
Electron gyrofrequency & $\omega_c$	& $\displaystyle eB/m_e$ s$^{-1}$	& $eB/m_e$ s$^{-1}$\\
Electron-ion collision frequency & $\nu_{ei}$ & $\displaystyle \frac{\pi^\frac{3}{2}n_e Ze^4\ln\Lambda}{2^\frac{1}{2}(4\pi\varepsilon_0)^2m_e^2v_{te}^3}$ s$^{-1}$&  $\displaystyle \frac{4(2\pi)^\frac{1}{2}n_e Ze^4\ln\Lambda}{3m_e^2v_{te}^3}$ s$^{-1}$\\
Coulomb-logarithm & $\ln\Lambda$	& $\displaystyle\ln \frac{9N_D}{Z}$ &$\displaystyle\ln \frac{9N_D}{Z}$\\
\hline\hline
\end{tabular}
\end{center}
\end{table}

\end{document}

%% file: macros.tex
\def\vep0{\varepsilon_0}
\def\eps0{\varepsilon_0}

\def\be{\begin{equation}}
\def\ee{\end{equation}}
\def\bi{\begin{enumerate}}
\def\ei{\end{enumerate}}
\def\bes{\[}
\def\ees{\]}
\def\bea{\begin{eqnarray}}
\def\eea{\end{eqnarray}}
\def\bar{\begin{eqnarray*}}
\def\ear{\end{eqnarray*}}

\newcommand{\dbyd}[2]{\frac{\partial #1}{\partial #2}}

\newcommand{\ddt}[1]{\frac{d #1}{d t}}
\newcommand{\curl}{\nabla\bm{\times}}
\newcommand{\half}{{1/2}}

\newcommand{\fhalf}{\frac{1}{2}}

\newcommand{\Wcm}{~Wcm$^{-2}$}

\newcommand{\cmcub}{~cm$^{-3}$}

\newcommand{\mum}{~$\mu$m}
\newcommand{\cross}{\bm{\times}}
\newcommand{\crossb}{\bm{\times}}

\newcommand{\eq}[1]{Eq.~(\ref{#1})}

\newcommand{\downbox}[1]{_{\mbox{\scriptsize\rm #1}}}
